\documentstyle[11pt]{article}
\begin{document}
\begin{titlepage}
\begin{flushright}
BRX TH-315 \\
HEP-TH/9303039 \\
5 March 1993
\end{flushright}
\vspace{1cm}
\begin{center}
{\large Division Algebras, (1,9)-Space-time, Matter-antimatter Mixing} \\
\vspace{1cm}
{\large Geoffrey Dixon} \\
Department of Physics \\
Brandeis University \\
Waltham, MA 02254 \\
\vspace{1cm}
and \\
\vspace{1cm}
Department of Mathematics \\
University of Massachusetts \\
Boston, MA 02125 \\
\vspace{1cm}
{\bf Abstract}
\end{center}
The tensor product of the division algebras, which is a kernel for
the structure of the Standard Model, is also a root for the Clifford
algebra of (1,9)-space-time.  A conventional Dirac Lagrangian, employing
the (1,9)-Dirac operator acting on the Standard Model hyperfield, gives
rise to matter into antimatter transitions not mediated by any gauge
field.  These transitions are eliminated by restricting the
dependencies of the components of the hyperfield on the extra six
dimensions, which appear in this context as a complex triple.
\vspace{1cm}
\normalsize
\footnoterule
\noindent
{\footnotesize e-mail: DIXON@BINAH.CC.BRANDEIS.EDU}
\end{titlepage}
\newpage

This article is an extension of my work on applying the tensor product of the
division algebras to the lepto-quark Standard Model [1-4] and beyond.  Although
it is selfcontained, many results derived previously are not rederived here.

Applications of the division algebras to particle physics [5-10] are not new,
nor are all the same.  This application, to the best of my knowledge, while
owing a debt to the work of G{\" u}rsey and G{\" u}naydin, is the only
one of its kind.  Like all applications of these algebras, however, it is
motivated by the attractive notion that the special structures of mathematics
play a role in the design of reality.  Most theorists share a faith - or at
least a hope - of this sort; here it has been allowed to become a guiding
principle.

In this article I present the first radical extension of my ideas beyond the
Standard Model and its foundation.  Because it combines the Standard Model with
(1,9)-space-time ({\bf R}$^{1,9}$), it may well prove a step toward the
development of a connection to, and a narrowing of, string theory, the
initial euphoria to which has - in the fashion of GUTs and SUSY - succumbed to
the curse of multiple realities.

The nontrivial real division algebras with unity are the complexes, {\bf C},
quaternions, {\bf Q}, and octonions, {\bf O}.  They are 2-, 4-, and
8-dimensional.  Multiplication tables for {\bf Q} and {\bf O} are
constructable from the following elegant rules:
\begin{equation}
\begin{array}{ccc}
\begin{array}{c} \mbox{Division} \\
\mbox{Algebra} \\ \end{array} & {\bf Q} & {\bf O} \\ \\
\begin{array}{c} \mbox{Imaginary}  \\
\mbox{Units} \end{array} & q_{i}, 1=1,2,3, & e_{a}, a=1,...,7, \\ \\
\begin{array}{c} \mbox{Anti-}  \\
\mbox{commutators} \end{array} & q_{i}q_{j}+q_{j}q_{i} = 2\delta_{ij}, &
                        e_{a}e_{b}+e_{b}e_{a} = 2\delta_{ab}, \\ \\
\begin{array}{c} \mbox{Cyclic}  \\
\mbox{Rules} \end{array} &  q_{i}q_{i+1}=q_{i-1}=q_{i+2}, &
                      e_{a}e_{a+1}=e_{a-2}=e_{a+5}, \\ \\
\begin{array}{c} \mbox{Index} \\
                 \mbox{Doubling} \\ \end{array} &
\begin{array}{c} q_{i}q_{j}=q_{k}\Longrightarrow \\
                         q_{(2i)}q_{(2j)}=-q_{(2k)}, \\
                        \end{array} &
                        \begin{array}{c}  e_{a}e_{b}=e_{c}\Longrightarrow \\
                         e_{(2a)}e_{(2b)}=e_{(2c)}, \\
                        \end{array} \\

\end{array}
\end{equation}
where {\bf Q}-indices run from 1 to 3, modulo 3, and {\bf O}-indices run from 1
to 7,
modulo 7.

{\bf C}$\otimes${\bf Q} is spanned by the 8 elements $\{1, i, q_{j}, iq_{j}\}$.
It is isomorphic to the Pauli algebra, ${\bf C}(2)$, which is the Clifford
algebra of ${\bf R}^{3,0}$ space.  Represented by ${\bf C}(2)$, the spinors of
that Clifford algebra are $2\times 1$ over {\bf C}, the so-called Pauli or
Weyl spinors.  The spinor space of {\bf C}$\otimes${\bf Q}, however, is
$1\times 1$ over {\bf C}$\otimes${\bf Q}, hence is {\bf C}$\otimes${\bf Q}
itself.  In this case, to distinguish the Clifford algebra from its spinor
space, we denote the former ${\bf C}_{L}\otimes{\bf Q}_{L}$, the subscript
indicating action from the left on the spinor space, which we denote
{\bf C}$\otimes${\bf Q}.

{\bf C}$\otimes${\bf Q} is twice as large as it needs to be.  It is the direct
sum of two 2-dimensional (over {\bf C}) Weyl spinor spaces unmixed by
${\bf C}_{L}\otimes{\bf Q}_{L}$ (just as
$\left[\begin{array}{cc} x_{1} & y_{1} \\ x_{2} & y_{2} \\ \end{array}\right]$
in ${\bf C}(2)$ is the direct sum of the Weyl spinor spaces
$\left[\begin{array}{cc} x_{1} & 0 \\ x_{2} & 0 \\ \end{array}\right]$ and
$\left[\begin{array}{cc} 0 & y_{1} \\ 0 & y_{2} \\ \end{array}\right]$).
If $\vec{x}\in{\bf Q}$ satisfies $\vec{x}^{2} = -1$, then
multiplication from the right on {\bf C}$\otimes${\bf Q} by the idempotents
$\frac{1}{2}(1 \pm i\vec{x})$ projects two such Weyl spinor spaces
(just as multiplication from the right by the idempotents
$\frac{1}{2}(\left[\begin{array}{cc} 1 & 0 \\ 0 & 1 \\ \end{array}\right] \pm
\left[\begin{array}{cc} 1 & 0 \\ 0 & -1 \\ \end{array}\right])$ on
$\left[\begin{array}{cc} x_{1} & y_{1} \\ x_{2} & y_{2} \\ \end{array}\right]$
projects the ${\bf C}(2)$ Weyl spinor spaces above).
${\bf Q}_{R}$, which acts from the right on {\bf C}$\otimes${\bf Q}, mixes
these two independent spinor spaces.  ${\bf Q}_{R}$ commutes with
${\bf C}_{L}\otimes{\bf Q}_{L}$, so it is an "internal" algebra, where the
Clifford (geometric) algebra is "external".  The elements of unit length of
${\bf Q}_{R}$ form the group $SU(2)$, which in previous work along these lines
was manifested as the isospin gauge symmetry [1].

The octonion algebra is generally considered ill-suited to Clifford algebra
theory because {\bf O} is nonassociative, and Clifford algebras are
associative.  This problem disappears once we identify {\bf O} as the spinor
space of ${\bf O}_{L}$, the adjoint algebra of actions of {\bf O} on itself
from the left. ${\bf O}_{L}$ {\it is} associative.  ${\bf O}_{L}$ is linear in
actions of the form
\begin{equation}
e_{Lab...c}[x] = e_{a}(e_{b}(...(e_{c}x)...)),
\end{equation}
$x\in{\bf O}$.  For example, although $e_{1}e_{2} = e_{6}$,
$$
e_{L12}[x] = e_{1}(e_{2}x) \ne e_{6}x = e_{L6}[x]
$$
in general; and although $e_{1}(e_{2}e_{4}) = e_{7}$,
$$
e_{L124}[x] = e_{1}(e_{2}(e_{4}x)) \ne e_{7}x = e_{L7}[x]
$$
in general.  These are consequences of nonassociativity.
The elements $e_{Lab...c}$ satisfy
\begin{eqnarray}
& e_{Labcc...d} = -e_{Lab...d}, \nonumber \\
& e_{Lab...c} = \pm e_{Lpq...r},
\end{eqnarray}
pq...r an even-odd permutation of ab...c, and
\begin{equation}
e_{Lab...c}e_{Ldf...g} = e_{Lab...cdf...g}. \nonumber
\end{equation}
It is also not difficult to prove that $e_{L7654321}[x] = x$ for all $x$ in
{\bf O}.  Therefore, for example, using (4) and (5) one can easily prove
\begin{equation}
e_{L4567} = e_{L4567}e_{L7654321} = e_{L321}.  \nonumber
\end{equation}
That is, any element of {\bf O}$_{L}$ with four or more indices can be reduced
to an element with three indices or less.  So a complete basis for
{\bf O}$_{L}$ consists of the elements
\begin{equation}
1,\; e_{La}, \; e_{Lab}, \; e_{Labc}.
\end{equation}
Therefore {\bf O}$_{L}$ is 1+7+21+35=64-dimensional, and
{\bf O}$_{L} \simeq$ {\bf R}(8).  The embedding of parentheses in the
definition (2), implying (4),
trivially implies {\bf O}$_{L}$ is associative.

{\bf O}$_{L}$ is isomorphic to the Clifford algebra of the space
${\bf R}^{0,6}$, the spinor space of which is 8-dimensional over {\bf R}.
In this case the spinor space is {\bf O} itself, the object space of
{\bf O}$_{L}$.  It is significant that the dimensionality of {\bf O}
is correct in this case.  This is tied to the remarkable fact that the
algebra {\bf O}$_{R}$ of right adjoint actions of {\bf O} on itself is
the {\it same} algebra as {\bf O}$_{L}$.  Every action in {\bf O}$_{R}$
can be written as an action in {\bf O}$_{L}$.

A 1-vector basis for {\bf O}$_{L}$, playing the role of the Clifford algebra
of ${\bf R}^{0,6}$, is $\{e_{Lp}, p=1,...,6\}$.  The resulting 2-vector
basis is then $\{e_{Lpq}, p,q=1,...,6, p\ne q\}$.  This subspace is
15-dimensional, closes under the commutator product, and is in that case
isomorphic to $so(6)$.  The intersection of this Lie algebra with the Lie
algebra of the automorphism group of {\bf O}, $G_{2}$, is $su(3)$, with a
basis
\begin{equation}
su(3)\rightarrow \{e_{Lpq}-e_{Lrs}, p,q,r,s \mbox{ distinct, and from }
1\mbox{ to } 6\}.
\end{equation}
The group $SU(3)$ generated by these elements arises as the color gauge group
in applications [1] (note that $SU(3)$ is the stability group of $e_{7}$,
hence the index doubling automorphism of {\bf O} is an $SU(3)$ rotation).

Finally we let ${\bf C}\otimes{\bf Q}\otimes{\bf O}$ play the role of spinor
space to ${\bf C}_{L}\otimes{\bf Q}_{L}\otimes{\bf O}_{L}$, which is
isomorphic to {\bf C}(16), hence isomorphic to the Clifford algebra of the
space ${\bf R}^{0,9}$.  With respect to the gauge symmetry
$SU(2)\times SU(3)$ outlined above, which expands to $U(2)\times U(3)$ [1], the
spinor
space ${\bf C}\otimes{\bf Q}\otimes{\bf O}$ transforms exactly like the
direct sum of a family and antifamily of lepto-quark Weyl spinors.
Quantum numbers for the (family) spinors can be manifested in two ways,
one corresponding to righthanded particles, one to lefthanded.  They can
be simultaneously incorporated by expanding
${\bf C}_{L}\otimes{\bf Q}_{L}\otimes{\bf O}_{L}$ to
${\bf C}_{L}\otimes{\bf Q}_{L}\otimes{\bf O}_{L}(2)$ ($2\times 2$ over
${\bf C}_{L}\otimes{\bf Q}_{L}\otimes{\bf O}_{L}$), the "Dirac" algebra
for ${\bf R}^{1,9}$ space-time (just as ${\bf C}_{L}\otimes{\bf Q}_{L}(2)$,
isomorphic to ${\bf C}(4)$, is the Dirac algebra for ${\bf R}^{1,3}$).
The spinor space in this case is $2\times 1$ over
${\bf C}\otimes{\bf Q}\otimes{\bf O}$.

Let $\Psi$ be such a spinor, and give it a functional dependence on
${\bf R}^{1,9}$ space-time.  Let
\begin{equation}
\rho_{\pm}=(1 \pm ie_{7})/2.
\end{equation}
Then $\rho_{+}\Psi$ is the matter half of $\Psi$, and $\rho_{-}\Psi$ the
antimatter half.  $\rho_{+}\Psi\rho_{+}$ is an $SU(2)$ lepton doublet, and
$\rho_{+}\Psi\rho_{-}$ is a quark $SU(2)$ doublet, $SU(3)$ triplet
(reverse signs for antimatter).

Define in {\bf R}(2): \\
$$\epsilon = \left(\matrix{1&0\cr 0&1\cr}\right), \alpha =
\left(\matrix{1&0\cr 0&-1\cr}\right), \beta = \left(\matrix{0&1\cr
1&0\cr}\right), \omega = \left(\matrix{0&1\cr -1&0\cr}\right).$$ \\
A 1-vector basis for the Clifford algebra of $R^{1,9}$ consists of the
elements:
\begin{equation}
\gamma_{0}=\beta,\;\; \gamma_{j}=q_{j}e_{L7}\omega, j=1,2,3, \;\;
\gamma_{h}=ie_{h-3}\omega, h=4,...,9.
\end{equation}
These satisfy:
$$
\gamma_{h}\gamma_{l}+\gamma_{l}\gamma_{h} = 2\eta_{hl}\epsilon, \nonumber
$$
$\eta_{hl}$ diagonal ($1(+),9(-)$).

The (1,9)-Dirac operator is ${\not\!\partial}_{1,9}=\gamma_{f}\partial^{f},
f=0,1,...,9$,
and I define $\not\!\partial_{1,3}= \gamma_{\mu}\partial^{\mu}, \mu=0,1,2,3$,
$\not\!\partial_{0,6}=\not\!\partial_{1,9}-\not\!\partial_{1,3}$.  Define
\begin{equation}
\rho_{L\pm}=(1 \pm ie_{L7})/2
\end{equation}
(the left adjoint version of $\rho_{\pm}$).  Using these adjoint idempotents
we can decompose ${\not\!\partial}_{1,9}$ into its (1,3)- and (0,6)-Dirac
operator parts, one of each for both matter and antimatter: \newpage
$$
{\not\!\partial}_{1,9}  = \rho_{L+}{\not\!\partial}_{1,9}\rho_{L+}
                        + \rho_{L-}{\not\!\partial}_{1,9}\rho_{L-}
                        + \rho_{L+}{\not\!\partial}_{1,9}\rho_{L-}
                        + \rho_{L-}{\not\!\partial}_{1,9}\rho_{L+}
$$
$$
                        = \rho_{L+}{\not\!\partial}_{1,3}\rho_{L+}
                        + \rho_{L-}{\not\!\partial}_{1,3}\rho_{L-}
                        + \rho_{L+}{\not\!\partial}_{0,6}\rho_{L-}
                        + \rho_{L-}{\not\!\partial}_{0,6}\rho_{L+},
$$
\begin{equation}
                        = {\not\!\partial}_{1,3}\rho_{L+}
                        + {\not\!\partial}_{1,3}\rho_{L-}
                        + {\not\!\partial}_{0,6}\rho_{L-}
                        + {\not\!\partial}_{0,6}\rho_{L+}
\end{equation}
(note that ${\not\!\partial}_{1,3}\rho_{L\pm}$ are the matter/antimatter
Dirac operators for (1,3)-space-time, and that because $e_{L7}\rho_{L\pm}
=\mp i\rho_{L\pm}$, the partials of the latter are space-reflected
relative to the former).  Therefore,
$$
{\not\!\partial}_{1,9}\Psi
                        = ({\not\!\partial}_{1,3}\rho_{L+}
                        + {\not\!\partial}_{1,3}\rho_{L-}
                        + {\not\!\partial}_{0,6}\rho_{L-}
                        + {\not\!\partial}_{0,6}\rho_{L+})\Psi
$$
\begin{equation}
                        = {\not\!\partial}_{1,3}(\rho_{+}\Psi)
                        + {\not\!\partial}_{1,3}(\rho_{-}\Psi)
                        + {\not\!\partial}_{0,6}(\rho_{-}\Psi)
                        + {\not\!\partial}_{0,6}(\rho_{+}\Psi).
\end{equation}

To form a Lagrangian for the field we use the inner product of
${\bf C}\otimes{\bf Q}\otimes{\bf O}$ [1]:
\begin{equation}
\begin{array}{c}
{\cal L}=<\Psi,\not\!\partial_{1,9}\Psi> \\
        =<\rho_{+}\Psi+\rho_{-}\Psi,
                          {\not\!\partial}_{1,3}(\rho_{+}\Psi)
                        + {\not\!\partial}_{1,3}(\rho_{-}\Psi)
                        + {\not\!\partial}_{0,6}(\rho_{-}\Psi)
                        + {\not\!\partial}_{0,6}(\rho_{+}\Psi)> \\
        =<\rho_{+}\Psi, {\not\!\partial}_{1,3}(\rho_{+}\Psi)>
        +<\rho_{-}\Psi, {\not\!\partial}_{1,3}(\rho_{-}\Psi)> \\
        +<\rho_{+}\Psi, {\not\!\partial}_{0,6}(\rho_{-}\Psi)>
        +<\rho_{-}\Psi, {\not\!\partial}_{0,6}(\rho_{+}\Psi)> \\
\end{array}
\end{equation}
(the last equality arising from the algebra of the inner product).
The first two terms after the last equality in (13),
$<\rho_{\pm}\Psi, {\not\!\partial}_{1,3}(\rho_{\pm}\Psi)>$,
are ordinary.   One can obtain a list of viable particle transitions
from such Lagrangians, as each Weyl component of $\Psi$ has an
obvious particle identification.  For example, after gauging
$U(2)\times U(3)$, algebraic combinations of spinor and gauge fields
that survive the inner product correspond to viable transitions
(this aspect won't be developed further here; see [1],[11]).  These
first two terms connect
matter/antimatter to matter/antimatter ($\rho_{\pm}\Psi$=
matter/antimatter), and upon gauging $U(2)\times U(3)$ give rise to
an unconventional looking version of the Standard Model.

The last two terms of (13),
$<\rho_{\mp}\Psi, {\not\!\partial}_{0,6}(\rho_{\pm}\Psi)>$, are a
problem, even without gauge fields, for they imply matter/antimatter
($\rho_{\pm}\Psi$) into antimatter/matter ($\rho_{\mp}\Psi$)
transitions, mediated algebraically by ${\not\!\partial}_{0,6}$.
As such transitions are unobserved, the rest of the article will be
devoted to getting rid of the last two terms of (13).

The 2-vector basis for the Clifford algebra of ${\bf R}^{1,9}$, derived
from the 1-vectors in (8), is
\begin{equation}
q_{j}\epsilon, \; q_{j}e_{L7}\alpha, \; ie_{Lp}\alpha, \;
iq_{j}e_{Lp7}\epsilon, \; e_{Lpq}\epsilon,
\end{equation}
j=1,2,3, p,q$\in\{1,...,6\}$.  This 45-dimensional subspace closes under
the commutator product and is in that case isomorphic to $so(1,9)$.
The first six elements, $\{q_{j}\epsilon, \; q_{j}e_{L7}\alpha\}$,
form a basis for $so(1,3)$, the last fifteen, $\{e_{Lpq}\epsilon\}$,
a basis for $so(6)$.  This is the same $so(6)$ we saw earlier, and it
contains color $su(3)$ (see (7)).  That is, the space ${\bf R}^{0,6}$,
hence ${\not\!\partial}_{0,6}$, carry color charges (one consequence of
these charges: in none of the unwanted transitions implied by (13)
can a particle make a transition to its own antiparticle; hence, for
example, quarks may mix with anitleptons, violating baryon and lepton
number conservation).

Consider the element ${\not\!\partial}_{0,6}(\rho_{+}\Psi)$ which appears
in the last term of (13).  Because
\begin{equation}
\rho_{\pm}e_{7}=\mp i\rho_{\pm}, \; \rho_{\pm}e_{5}=\mp i\rho_{\pm}e_{1}, \;
\rho_{\pm}e_{3}=\mp i\rho_{\pm}e_{2}, \; \rho_{\pm}e_{6}=\mp i\rho_{\pm}e_{4},
\end{equation}
$\rho_{+}\Psi$ may be decomposed into
\begin{equation}
\rho_{+}\Psi = \rho_{+}[\Psi_{+}^{0} + \Psi_{+}^{1}e_{1}
                       + \Psi_{+}^{2}e_{2}+ \Psi_{+}^{4}e_{4}],
\end{equation}
where the $\Psi_{+}^{m}$, m=0,1,2,4, are $2\times 1$ over
${\bf C}\otimes{\bf Q}$.  These four
fields can be designated lepton, red-, green-, and blue-quark.

Now consider $ \not\!\partial_{0,6}(\rho_{+}\Psi_{+}) $, and
in particular, for example, the term (sum p=1,...,6)
\begin{eqnarray}
&& \not\!\partial_{0,6}(\rho_{+}\Psi_{+}^{1}e_{1})
=i\omega e_{p}\partial^{p+3}[\rho_{+}\Psi_{+}^{1}e_{1}] \nonumber \\
&& =i\omega(\rho_{-}e_{1}\partial^{4}+ \rho_{+}e_{2}\partial^{5}+
\rho_{+}e_{3}\partial^{6}+ \rho_{+}e_{4}\partial^{7}+ \rho_{-}e_{5}\partial^{8}
+\rho_{+}e_{6}\partial^{9})[\Psi_{+}^{1}e_{1}] \nonumber \\
&& =i\omega(\rho_{-}e_{1}(\partial^{4}+i\partial^{8})+
\rho_{+}e_{2}(\partial^{5}-i\partial^{6})+
\rho_{+}e_{4}(\partial^{7}-i\partial^{9}))[\Psi_{+}^{1}e_{1}] \nonumber \\
&& =i\omega(e_{1}(\partial^{4}+i\partial^{8})+
e_{2}(\partial^{5}-i\partial^{6})+
e_{4}(\partial^{7}-i\partial^{9}))[\rho_{+}\Psi_{+}^{1}e_{1}] \nonumber \\
&& \equiv \not\!\partial_{6+--}[\rho_{+}\Psi_{+}^{1}e_{1}]
\end{eqnarray}
(in the second line the nonassociativity of {\bf O} plays a part
in  altering the sign subscripts of $\rho_{\pm}$; in general nonassociativity
plays an essential role in keeping the mathematics consistent with
phenomenology).
$ \not\!\partial_{6+--} $ (generalized below) is defined  in the penultimate
line.
In like manner one can demonstrate that
\begin{eqnarray}
& \not\!\partial_{6}\rho_{+}\Psi_{+}^{0} =
\not\!\partial_{6+++}\rho_{+}\Psi_{+}^{0},
\nonumber \\
 & \not\!\partial_{6}(\rho_{+}\Psi_{+}^{2}e_{2}) =
\not\!\partial_{6-+-}(\rho_{+}\Psi_{+}^{2}e_{2}) \nonumber \\
& \not\!\partial_{6}(\rho_{+}\Psi_{+}^{4}e_{4}) =
\not\!\partial_{6--+}(\rho_{+}\Psi_{+}^{4}e_{4})
\end{eqnarray}
(no parentheses are needed in the first of these equations (lepton term),
for nonassociativity only becomes an issue on the quark terms).
For any real variables x and y, and differentiable f:
$(\partial_{x}+i\partial_{y})f(x+iy) = 0$.  Therefore, ignoring {\bf R}$^{1,3}$
coordinates, if
\begin{eqnarray}
& \Psi_{+}^{0} = \Psi_{+}^{0}(x_{4}+ix_{8},x_{5}+ix_{6},x_{7}+ix_{9}),
\nonumber \\
& \Psi_{+}^{1} = \Psi_{+}^{1}(x_{4}+ix_{8},x_{5}-ix_{6},x_{7}-ix_{9}),
\nonumber \\
& \Psi_{+}^{2} = \Psi_{+}^{2}(x_{4}-ix_{8},x_{5}+ix_{6},x_{7}-ix_{9}),
\nonumber \\
& \Psi_{+}^{4} = \Psi_{+}^{4}(x_{4}-ix_{8},x_{5}-ix_{6},x_{7}+ix_{9}),
\end{eqnarray}
then
\begin{equation}
\not\!\partial_{0,6}(\rho_{+}\Psi) = 0.
\end{equation}

The antimatter fields of $\rho_{-}\Psi$ would have functional dependencies
conjugate to those above.  Any fluctuation from these would give rise to
unobserved matter-antimatter mixing.

Under $U(3)$ the lepton term $\Psi_{+}^{0}$ is supposed invariant, but its
3 complex coordinates in (19) are not.  In making $U(3)$ a local gauge
symmetry, dependent upon ${\bf R}^{1,3}$ coordinates, the complex
coordinates of $\Psi_{+}^{0}$ also acquire a functional dependence on
${\bf R}^{1,3}$.  The orbit of $U(3)$ is $S^{5}$, the 5-sphere.  Because
$\Psi_{+}^{0}$ is dependent on 3 complex coordinates, and not 6 real,
this precludes a variation of $\Psi_{+}^{0}$ by even so much as a phase
factor under $U(3)$.  It would seem then that the colorless lepton term
$\Psi_{+}^{0}$ must be independent entirely of the color-carrying
coordinates of ${\bf R}^{0,6}$.

The complex triple associated with $\Psi_{+}^{1}$ in (19) has a more
complicated SU(3) transformation, further complicated by the fact that
$\Psi_{+}^{1}$ is itself simultaneously transformed.  However, $\Psi_{+}^{1}$
is invariant under the action of the SU(2) subgroup
of SU(3) that leaves $e_{1}$ and $e_{5}$ invariant.  Following the same
reasoning used above we now conclude that $\Psi_{+}^{1}$ must be independent,
not of all of {\bf R}$^{0,6}$ as was $\Psi_{+}^{0}$, but of $x_{r}$, r=5,6,7,9.

In general we may now conclude, inorder to preserve (20), that
\begin{eqnarray}
& \Psi_{+}^{0} = \Psi_{+}^{0}(x_{\mu},...,...,...),
\nonumber \\
& \Psi_{+}^{1} = \Psi_{+}^{1}(x_{\mu},x_{4}+ix_{8},...,...),
\nonumber \\
& \Psi_{+}^{2} = \Psi_{+}^{2}(x_{\mu},...,x_{5}+ix_{6},...),
\nonumber \\
& \Psi_{+}^{4} = \Psi_{+}^{4}(x_{\mu},...,...,x_{7}+ix_{9}),
\end{eqnarray}
where (,...) indicates independence of the complex coordinate in that slot, and
$x_{\mu}$ denote the coordinates of {\bf R}$^{1,3}$.

Does any of this have anything to do with string theory?  I confess myself not
a string theorist, so I can not supply a definitive answer to that question.
String theory uses ${\bf R}^{1,9}$, and it deals with the extra 6 dimensions
by balling them up into a complex 3-manifold
too small to be observed.  My route to ${\bf R}^{1,9}$ is
certainly different, but in requiring (20) the space ${\bf R}^{0,6}$ is
forced to appear in the guise of a complex 3-space.  It has not yet been
investigated if some specific compactification is required of the model, much
less if there is an associated $SU(3)$ holonomy group [13].  As to its
unobservability, everything in this model (specifically quarks and
${\bf R}^{0,6}$) associated with the octonion units $e_{p}, p=1,...,6$
(also associated with nonassociativity) is unobserved.  There may be some nice
algebraic/quantum mechanical explanation for this, but even so one finds such
subtlety is generally manifested by more prosaic explanations as well, like
infrared slavery, and, presumably, compactification. \\  \\

References:

1. G.M. Dixon:  {\sl Il Nuovo Cim} {\bf 105B} 349(1990).

2. G.M. Dixon:  {\sl Phys. Lett.} {\bf 152B} 343(1985).

3. G.M. Dixon:  {\sl Phys. Rev.} D {\bf 29} 1276(1984).

4. G.M. Dixon:  {\sl Phys. Rev.} D {\bf 28} 833(1983).

5. M. Gunaydin and F. Gursey: {\sl Phys. Rev.} D {\bf 10} 674(1974).

6. J. Lukierski and P. Minnaert: {\sl Phys. Lett.} {\bf 129B} 392(1983).

7. R. Foot and G.C. Joshi: {\sl Phys. Lett.} {\bf 199B} 203(1987).

8. K-W. Chung and A. Sudbery: {\sl Phys. Lett.} {\bf 198B} 16(1987).

9. P. Goddard, W. Nahm, D.I. Olive, H. Ruegg and A. Schwimmer: {\sl Comm. Math.
Phys.} {\bf 112}, 385(1987).

10. J.A. Harvey and A. Strominger: {\sl Phys. Rev. Lett.} {\bf 66}, 549(1991).

11. G.M. Dixon: monograph in preparation.

12. P. Candelas, G.T. Horowitz, A. Strominger, E. Witten,
{\it Superstring Phenomenology}, "Symposium on Anomolies, Geometry,
Topology", World Scientific Publishing Co. Pte. Ltd., 377(1985).

Acknowledgement:  I would like to thank Prof. Hugh Pendleton of Brandeis
University for helpful discussions.

\end{document}